\documentclass{llncs}
\pdfoutput=1

\usepackage{enumitem,xcolor}

\newcounter{myprot}
\newcounter{mythm}
\newenvironment{mythm}
{\refstepcounter{mythm} \vspace{.5em} \noindent \textbf{THEOREM \arabic{mythm}:}}
{\vspace{.5em}}

\newcounter{mycor}
\newcounter{myobs}
\newcounter{mydef}
\newenvironment{mydef}
{\refstepcounter{mydef} \vspace{1em} \noindent \textbf{DEFINITION \arabic{mydef}:}}
{\vspace{.5em}}

\newcounter{myconj}
\newenvironment{myproof}
{\noindent \textbf{PROOF:}}
{\vspace{-2em}\begin{flushright} 
$\qed$ 
\end{flushright}}

\usepackage[font=sf]{caption}
\usepackage{amsmath,mathtools}
\usepackage[hyphens]{url}
\usepackage{xfrac}
\usepackage{wrapfig,multirow,relsize}
\usepackage[nospace,compress]{cite}
\usepackage{microtype} 

\usepackage{xspace}

\newcommand{\para }[1]{\smallskip \noindent {\bf #1}}

\usepackage{algorithmic}
\usepackage{booktabs} 

\usepackage[
rm={oldstyle=false,proportional=true},
sf={oldstyle=false,proportional=true},
tt={oldstyle=false,proportional=true,variable=true},
qt=false
]{cfr-lm}
\makeatletter
\g@addto@macro{\UrlBreaks}{\UrlOrds}
\makeatother

\makeatletter
\def\@copyrightspace{\relax}
\makeatother

\title{Securing the Assets of Decentralized Applications using Financial Derivatives (DRAFT)}
\author{George Bissias$^\dagger$, Brian  Levine$^\dagger$, Nikunj Kapadia$^\circ$ \\
$\dagger$College of Information and Computer Sciences\\
$\circ$Isenberg School of Management\\
University of Massachusetts, Amherst}
\institute{}
\makeatletter
\def\blfootnote{\xdef\@thefnmark{}\@footnotetext}
\makeatother

\begin{document}

\maketitle
\urlstyle{sf}

\abovedisplayskip=3pt
\belowdisplayskip=3pt
\abovedisplayshortskip=0pt
\belowdisplayshortskip=4pt

\begin{abstract}
Ethereum contracts can be designed to function as fully decentralized
applications called DAPPs. Many DAPPs have already been fielded,
including an online marketplace, a role playing game, a prediction
market, and an Internet service provider.  Unfortunately, DAPPs can be
hacked, and the assets they control can be stolen. A recent attack on
an Ethereum decentralized application called The DAO demonstrated that
smart contract bugs are more than an academic concern. Ether worth
tens of millions of US dollars was extracted by an attacker from The
DAO, sending the value of its tokens and the overall exchange price of
ether tumbling.

We present a market-based technique for insuring the ether holdings of
a DAPP using futures contracts indexed by the trade price of ether for
DAPP tokens.  Under fairly general circumstances, our technique is
capable of recovering the majority of ether lost from theft with high
probability even when all of the ether holdings are stolen; and the
only cost to DAPP token holders is an adjustable ether withdrawal
fee. If the probability of a margin call in $d$ days is $p$ for a
futures contract with 20 times leverage, then our approach will allow
for the recovery of half the stolen ether with probability $p$ and a
withdrawal fee of 5\%. A higher withdrawal fee of 25\% allows for more
than 80\% of the ether to be recovered with probability $p$.
\end{abstract}

\section{Introduction}

Apart from its utility as a currency, one of the greatest features of Bitcoin~\cite{Nakamoto:2009} is its facility for creating cryptographically secure smart contracts. Some possible applications include a property ownership registry, micropayment platform, third-party escrow system, and cryptocurrency exchange~\cite{bitcoin:contracts}. However, Bitcoin provides only limited tools for smart contract creation. A newer platform called Ethereum\cite{ethereum} was built from the ground up as a general purpose computing system for smart contracts with much more sophisticated functionality. Ethereum contracts  can be designed so as to function as fully {\em decentralized applications} called \emph{DAPPs}. Many DAPPs have already been fielded, including an online marketplace~\cite{SafeMarket:2016}, a role playing game~\cite{etheria:2016}, a prediction market~\cite{Gnosis:2016}, and  an Internet service provider~\cite{oraclize:2016}. 

Unfortunately, DAPPs can be hacked, and the assets they control can be stolen. For example, in May 2016, a DAPP called  the  {\em Decentralized Autonomous Organization} (The DAO) was created as a type of decentralized hedge fund. It raised over US\$150,000,000 worth of ether (Ethereum's native cryptocurrency) during a crowd sale, where \emph{tokens} were issued (in tranches) at a fixed exchange rate for ether~\cite{Siegel:2016}. Token holders were allowed to submit proposals for how to invest The DAO's ether, and tokens also afforded the holders with voting rights on what proposals should be pursued. For example, one proposal was funding for a company called Slock.it~\cite{slock:2016} whose technology enables users to securely share any item, such as a bicycle or an apartment, with a third party. A key feature of The DAO was that ether could be withdrawn. Token holders were allowed to split off a \emph{child} DAO at any time, taking with them their share of the ether. After a provisional 28-day waiting period, the child DAO would be allowed to execute proposals itself, one which could be to withdraw its ether holdings.

 On June 17, 2016 an attacker began an unauthorized transfer of ether from The DAO into a new child DAO~\cite{Castillo:2016}. By June 18, more than US\$100,000,000 in ether had been stolen, locked in the child DAO for the 28-day waiting period, after which the attacker could take possession~\cite{Siegel:2016}. The Ethereum community struggled over how to handle the theft. Ultimately, the community was literally divided when the majority of the Ethereum developers decided to rewrite the blockchain in order to return the ether to the original token holders, while a smaller portion continued to mine on the old blockchain~\cite{Castillo:2016a}. Even on the old blockchain, a \emph{white hat} group of hackers were able to steal back a large portion of the stolen ether~\cite{Torpey:2016}. Nevertheless, the hack led to the dissolution of The DAO, split the community, and caused the price of ether to drop by approximately two thirds.

\para{Contribution.}
In this paper, we present a market-based technique for insuring the ether holdings of a DAPP using futures contracts indexed by the trade price of ether for DAPP tokens. Note that we are not proposing a method of actually securing contract code, but rather we are developing a process for protecting a DAPP's assets in the (perhaps inevitable) event that a software security flaw is discovered and exploited. Under fairly general circumstances, \emph{our technique is capable of recovering the majority of ether lost from theft with high probability even when all of the ether holdings are stolen}; and the only cost to DAPP token holders is an adjustable ether withdrawal fee. The probability of successful recovery is tied to the volatility of the futures contracts. For example, suppose that it takes as many as $d$ days for investors to become aware of a DAPP theft where all ether are stolen. If the probability of a margin call in $d$ days is $p$ for a futures contract with 20 times leverage, then our approach will allow for the recovery of half the stolen ether with probability $p$ and a withdrawal fee of 5\%. A higher withdrawal fee of 25\% allows for more than 80\% of the ether to be recovered with probability $p$.

\section{Problem Statement}
\label{sec:problem}
 
Our goal is to secure the financial assets held in DAPPs like The DAO despite successful exploitation of a vulnerabilities in their contract code. Although it may be too difficult to absolutely determine the security of a contract's software, it still may be possible to secure the contract's assets using financial derivatives as a form of insurance. 

To illustrate our approach, consider a DAPP $\mathcal{D}$ that holds an initial funding period in which $n$ total tokens (TOK) are sold in exchange for 1 ether (ETH) each. Similarly, in exchange for 1 TOK, $\mathcal{D}$ permits the withdrawal of ETH proportional to the ratio of total ETH holdings to outstanding TOK. 
TOK holders are given voting rights in $\mathcal{D}$, which allows them to vote on how the ETH held in the contract are spent; they can either vote to invest the ETH in a third-party, or directly withdraw the ETH in exchange for their TOK. Suppose that there are $k$ TOK outstanding and $j$ ETH remain in $\mathcal{D}$. Then, for the purpose of exposition, we define an \emph{authorized withdrawal} as one that trades $m$ TOK for $m(\sfrac{j}{k})$ ETH; any withdrawal that trades $m$ TOK for more ETH is considered \emph{unauthorized}. Note, however, that our approach does not require for withdrawals to be explicitly labeled as authorized or unauthorized.

\para{Basic assumptions.} We assume for simplicity that, prior to attack, all $n$ ETH remain in $\mathcal{D}$ and are available for withdrawal. The attacker is assumed to be capable of conducting an unauthorized withdrawal of $m$ ETH without sacrificing any TOK. The DAO tokens traded on an open market and carried an exchange price denominated in ETH. Thus, we assume that TOK can also be traded for ETH. Throughout this document, we refer to the exchange price of ETH for TOK, denoted ETH/TOK (ether per token). We further assume that one of the existing futures exchanges~\cite{okcoin:2016,bitmex:2016} will implement \emph{futures contracts} for speculating on the exchange price ETH/TOK. We feel this is a reasonable assumption given that tokens are implemented as distinct cryptocurrencies and exchanges are quick to implement futures contracts for popular cryptocurrencies. Finally, we assume that, like ether, tokens cannot be stolen due to cryptographic protections. Rather, only the contract code is assumed to be at risk, since it is code specific to $\mathcal{D}$.

\para{DAPP Price Assumption.} We also make one more assumption, which is set apart here because it is critical to our approach: the exchange price ETH/TOK is equal to the ratio of ETH held by $\mathcal{D}$ to the number of TOK outstanding. Specifically, prior to attack the price of ETH/TOK is 1 and an unauthorized withdrawal of $m \leq n$ ETH will result in a drop to $\sfrac{m}{n}$. This assumption is justified if we assume that the trade of price of ETH/TOK is determined solely by the amount of ETH that can be redeemed for each TOK.

\section{Background and Related Work}

\subsection{Bitcoin}

Using a blockchain as a method for distributed consensus was first proposed by Nakamoto as part of a proposal for a cryptocurrency called Bitcoin~\cite{Nakamoto:2009}. Blockchains allow for an open group of peers to reach consensus, while mitigating Sybil attacks\cite{Douceur:2002} and the limitations of the FLP impossibility result~\cite{Fischer:1985}.  Unlike physical coins, bitcoins do not exist as distinct items, but rather as an account balance in an {\em address}, and therefore are fungible, divisible, and recombinable. Addresses comprise a stored asymmetric cryptographic key and an associated balance. Bitcoins are exchanged through {\em transactions}, which authorize the movement of bitcoin between addresses. A transaction is a message that is cryptographically signed by the payer, specifying the exchange of some coins from the payer's address to the recipient's address. They are broadcast over Bitcoin's peer-to-peer network. Miners on the network each independently agglomerate a set of transactions into a {\em block}, verify that the transactions are valid, and attempt to solve a predefined proof-of-work problem involving this block {\em and all prior valid blocks}. The first miner to solve the proof-of-work problem broadcasts her solution to the network, adding it to the ever-growing {\em blockchain}; the miners then start over, trying to add a new block containing the set of transactions that were not previously added. When transactions appear on a block, they are considered {\em confirmed} and are honored by all participants in subsequent transactions. 

\para{Smart-contracts in Bitcoin.} It is currently possible to implement a limited range of smart-contracts with Bitcoin using its stack-based scripting language. These scripts can be be attached to transactions and run as part of the transaction validation process. The language intentionally lacks high level features like loops in order to prevent an attacker from forcing miners to spend unbounded resources during validation. Thus it is not currently possible to implement general purpose smart-contracts in Bitcoin. However, the RootStock~\cite{rootstock:2016} project promises to provide Turing-complete smart-contracts by introducing a Bitcoin sidechain~\cite{Back:2014} with the necessary protocol modifications.

\subsection{Ethereum}

Strictly for explanatory purposes, Ethereum can be thought of as a system that starts from the mechanisms in Bitcoin but with native support for Turing-complete smart-contracts builtin. Ethereum's mining process incentivizes miners to execute the code that comprises contracts by rewarding them with \emph{ether}, a cryptocurrency that, like bitcoin, has an exchange rate with fiat currencies, such as the US dollar. In addition to regular accounts analogous to Bitcoin addresses, Ethereum implements \emph{contract accounts}, which hold a balance of ether and whose behavior is controlled by arbitrarily complex, Turing-complete code. This code can be triggered to execute when a user submits a transaction to the miners which specifies the contract to run and also provides the ether necessary to pay for computation. Because code in contracts can be arbitrarily complex, the real cost to miners for running and validating the code can also be arbitrarily high. To address this issue, Ethereum assigns a cost to every low-level unit of computation called an \emph{opcode} in terms of a new unit of currency called \emph{gas}. Common examples of opcodes include \texttt{ISZERO}, \texttt{XOR}, and \texttt{ADD}. When a user submits a transaction calling for contract execution, he also specifies the gas price (the exchange price of gas for ether) and includes ether sufficient to cover the total cost of computation. If a miner feels that the gas price quoted in the transaction is sufficient, then she will add the transaction to her latest block. The contract code is executed every time that block is validated, which means that it is executed at least once by every miner and by every full node in the Ethereum network.

\subsection{Smart-Contract Security}


Research on smart-contract security is nascent in the cryptocurrency community. Buterin~\cite{Buterin:2017} offers some general recommendations based on the types of attacks witnessed in Ethereum to-date. He advocates for the development of layered, incremental defenses, but also argues that \emph{identifying} an attack is fundamentally difficult. In particular, users with different values might have differing opinions on the legitimacy of any given withdrawal. 

\section{A DAPP Insurance Mechanism}

A major problem with ethereum contracts  is that the  logic can be very complex. Contract software can represent an ample opportunity for an attacker to exploit a vulnerability in the contract to conduct an unauthorized ETH withdraw. In this section, we introduce a mechanism that protects the ether held in a contract  \emph{even in the event that an attacker is capable of making an unauthorized withdrawal at no expense}. Our solution is for the contract to retain a withdrawal fee from the ether to be withdrawn that will be used to purchase insurance, in the form of futures contracts FUT that hedge against a decrease in the trade price of ETH/TOK. If the withdrawal was  unauthorized, then presumably the trade price will eventually drop. At this point the contract sells its FUT for ETH, recovering some portion of the stolen ETH. However, in order to prevent iterative attacks, the recovered ETH are not immediately made available for withdrawal. They are locked in a recovery cache until TOK holders vote for their release (after the withdrawal code has been patched). 

\begin{figure}[!tbp]
  \centering
  \begin{minipage}[b]{1.0\textwidth}
    \includegraphics[width=\textwidth]{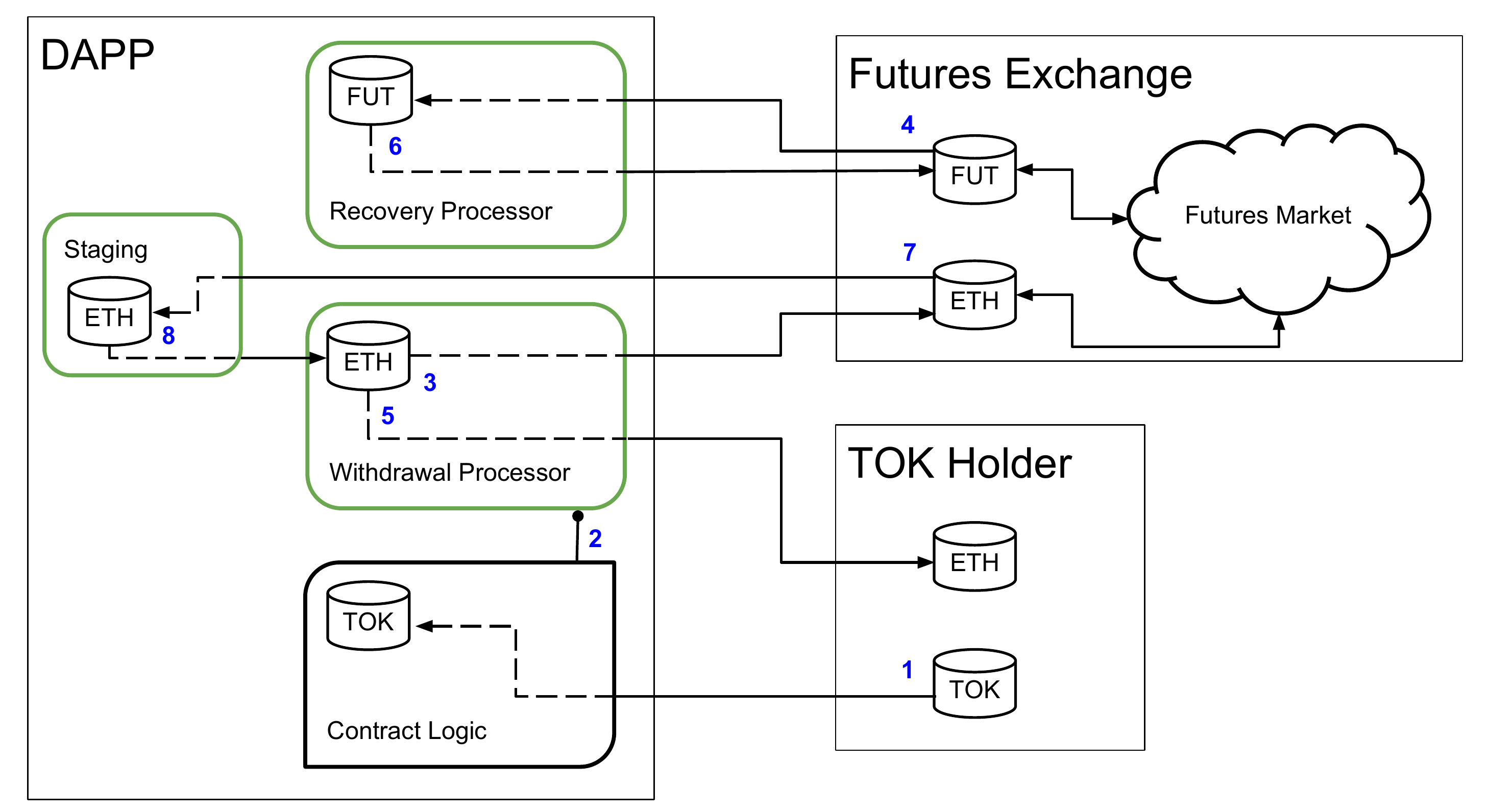}
    \caption{Insurance mechanism.}
    \label{fig:insurance_mech}
  \end{minipage}
\end{figure}

Figure~\ref{fig:insurance_mech} shows the withdrawal workflow. We have introduced three general purpose withdrawal components (shown in green) that operate independently from the existing contract-specific logic. The idea is that the code used to implement these components can be thoroughly tested and reused among all DAPPs that implement a withdrawal mechanism. The withdrawal process unfolds in eight steps (each delineated in blue in the figure), which we describe in detail below for a contract $\mathcal{D}$ as defined in  Section~\ref{sec:problem}.

\begin{enumerate}
\item A user interacts with the contract-specific logic of $\mathcal{D}$ in some way so as to initiate the withdrawal of ETH. In general, it is possible that a complex sequence of events unfold that lead to the withdrawal, but the most typical scenario is the deposit TOK. 
\item The contract logic signals its approval of the withdrawal of $m$ ETH to the withdrawal processor.
\item The withdrawal processor removes $m$ ETH from its cache and separates a portion $fm$ for the withdrawal fee. The fee is passed to an account controlled by $\mathcal{D}$ on the futures exchange. The $(1-f)m$ remaining ETH are momentarily held by the withdrawal processor.
\item The withdrawal processor trades $fm$ ETH on the futures exchange for $m$ futures contracts FUT that hedge against a drop in the trade price of ETH/TOK. The FUT are moved from an account controlled by $\mathcal{D}$ on the futures exchange to a cache controlled by the recovery processor.
\item The withdrawal processor releases $(1-f)m$ ETH to the user who initiated the withdrawal. This is the final responsibility of the withdrawal processor.
\item The recovery processor monitors the value of the FUT held in the account controlled by $\mathcal{D}$ on the futures exchange. Eventually the recovery processor trades the FUT for ETH.
\item Upon withdrawal from the futures exchange, the ETH are stored in a staging area that is intentionally quarantined from the rest of the contract logic.
\item Periodically, the current TOK holders vote to release the ETH in the staging area back into the main ETH cache controlled by the withdrawal processor. This manual voting process is used to ensure that an attacker cannot drain the ETH holdings of $\mathcal{D}$ multiple times before a potential exploit is discovered.
\end{enumerate}

\section{Mechanics of Futures Contracts}
\label{sec:back:futures}

Futures contracts, or simply \emph{futures}, are instruments that are created and brokered by a \emph{futures exchange}. They are an agreement between two parties, which are issued in pairs to individuals who enter opposing \emph{positions}. Central to the mechanics of futures contracts is the notion of the \emph{index price}, or the exchange price of ETH/TOK on some well-known currency exchange (typically independent from the futures exchange). Each contract is stipulated in terms of the \emph{settlement price}, which is the average index price during the hour prior to the \emph{settlement date}. We next define a particular type of futures contract that hedges against fluctuations in the exchange price of ETH/TOK.

\begin{mydef}
A \emph{futures contract} is an agreement between \emph{guarantor} and \emph{beneficiary} whereby the guarantor agrees to pay the beneficiary 1 TOK worth of ETH at the settlement price on the settlement date. 
\end{mydef}

\para{Positions.} A long position affords a trader with the \emph{right} to receive ETH at settlement while a short position encumbers a trader with the \emph{obligation} to deliver ETH at settlement. New futures are initially created directly by the exchange which matches pairs of traders willing to assume opposing positions. In return for agreeing to encumber himself with an SP, a trader receives 1 TOK worth of ETH at the current index price from the opposing trader entering an LP. In this case traders are said to have \emph{opened} a position.

\begin{mydef}
A trader opens a \emph{long position} (LP) or \emph{short position} (SP) whenever she becomes the beneficiary or guarantor, respectively, of a futures contract.
\end{mydef}

\para{Market.} An open position is automatically closed at settlement, but it is also possible for an agent to close a position by transferring her rights or obligations to a new trader. This transfer is conducted on the \emph{futures market}. In this market, a trader buys (sells) a futures contract in order to open a long (short) position, and he sells (buys) the contract in order to close a long (short) position. ``Buying'' is understood to mean either becoming the beneficiary of a contract (in the case of LPs) or being released from the obligation of a contract (in the case of SPs). Similarly, ``selling'' when entering an LP means giving up the right as beneficiary of the contract while it means becoming guarantor of a contract when entering an SP. The trade price or \emph{price} refers the amount of ETH a trader agrees to pay (receive) when buying (selling) a contract. When the futures market is efficient, futures will trade close to the current index price indicating that any information about the future settlement price is captured by the current index price. 

\para{Profit.} Profits and losses (P/L) associated with holding a position on a futures contract closely track the movement of the underlying index price.

\begin{mythm} 
\label{thm:lp_price}
Assuming efficient markets, a \emph{rise} (\emph{fall}) in the price of ETH/TOK by fraction $p$ while holding a long position will cause a corresponding \emph{profit} (\emph{loss}) of $p$ worth of ETH at the settlement date. 
\end{mythm}

\begin{myproof}
Suppose an LP is opened for $\varepsilon$ ETH in the futures market. Assuming the market is efficient, the trade price at that time would have been $\varepsilon$ ETH/TOK. If the price subsequently rises to $((1+p)\varepsilon)$ ETH/TOK, then the beneficiary claims $((1+p)\varepsilon)$ ETH at the time of settlement. Thus the trader in an LP has earned profit $((1+p)\varepsilon - \varepsilon)$ ETH = $p\varepsilon$ ETH. Similar reasoning can be used to show that a drop by fraction $p$ in price will lead to a commensurate loss of $p$ ETH for the trader.
\end{myproof}

\noindent Holding a short position has the opposite effect on P/L.

\begin{mythm} 
\label{thm:sp_price}
Assuming efficient markets, a \emph{rise} (\emph{fall}) in the price of ETH/TOK by fraction $p$ while holding a short position will cause an opposing \emph{loss} (\emph{profit}) of $p$ worth of ETH at the settlement date. 
\end{mythm}

\begin{myproof}
Suppose that an SP is opened such that the trader receives $\varepsilon$ ETH in exchange for encumbering himself as the guarantor of a futures contract. Again assuming that the market is efficient, the trade price would have been $\varepsilon$ ETH/TOK. If the price subsequently falls to $((1-p)\varepsilon)$ ETH/TOK, then the guarantor is obligated to pay the beneficiary just $((1-p)\varepsilon)$ ETH at the time of settlement. Thus the trader has earned profit $(\varepsilon - \varepsilon(1-p))$ ETH = $p\varepsilon$ ETH. Similar reasoning shows that a rise in index price by fraction $p$ will lead to an opposing loss of $p$ ETH for the trader.
\end{myproof} 

\noindent Notice that LPs have a finite potential downside of 100\% loss but an unbounded potential upside since there is no bound on the index price of ETH/TOK. On the other hand, SPs have an unbounded potential downside and a finite potential upside of 100\% gain, which occurs when the index price of ETH/TOK drops to zero.
  
\para{Margin.} We have shown that holding a position can result in either a profit or loss as a contract's trade price fluctuates. Any losses are payable when the position is closed or upon contract settlement. For each open position, a trader maintains \emph{margin} with the exchange, which is ETH collected at the time the position is opened for the express purpose of \emph{covering} any potential losses. In order to enforce the futures contract, the exchange ensures that every trader has enough margin to cover each open position in the event that it was closed at the current market price. When a trader lacks sufficient margin to cover her open position, the exchange initiates a \emph{margin call}: margin is confiscated by the exchange and used to close the open position. The trader loses both the margin and any right or obligations associated with the futures contract.
  
\para{Leverage.} The \emph{leverage} of a position refers to the percentage of margin required relative to a futures contract's purchase (sale) price at the time a long (short) position is opened. A $\lambda$-leveraged futures contract requires only fraction $1/\lambda$ TOK worth of ETH in margin, where typically $\lambda \in [1,100]$. 

\begin{mydef}
A $\lambda$-LP ($\lambda$-SP) is a $\lambda$-leveraged long (short) position that can be opened by posting $1 / \lambda$ TOK worth of ETH in margin at the current trade price.  
\end{mydef}

\section{Analysis of the Insurance Mechanism}

In this section we provide technical details and analysis for the withdrawal and recover processors. We introduce a theoretical framework for reasoning about the tradeoff between cost and security and provide concrete guidance on how to set security parameters.

\subsection{Withdrawal Processor}

$\mathcal{D}$ will implement a particular withdrawal mechanism such that any time $m$ ETH are withdrawn, it will reserve $\sfrac{\delta m}{\lambda}$ ETH in order to open $\delta m$ $\lambda$-SPs (see Section~\ref{sec:back:futures}). In this case $\mathcal{D}$ is said to have implemented a $(\delta, \lambda)$ \emph{insurance policy} and required a $\sfrac{\delta m}{\lambda}$ \emph{fee}. Suppose that initially there are $n$ TOK outstanding and $n$ ETH in $\mathcal{D}$. Now imagine that the $m$ ether were withdrawn illegitimately as a result of an attacker exploiting some vulnerability in $\mathcal{D}$ such that he is capable of completing the withdrawal without depositing any TOK. Immediately after the withdrawal, each TOK is redeemable for just $\sfrac{n-m}{n}$ ETH (this follows from the DAPP behavior defined in Section~\ref{sec:problem}). Thus, by the DAPP Price Assumption, the exchange price ETH/TOK will drop, but not all the way to $\sfrac{n-m}{n}$ because the $\lambda$-SPs still hold value, a fact which investors will be aware of. $\mathcal{D}$ will then close the $\lambda$-SPs, reclaiming some amount of the stolen ETH, during a \emph{recovery process}.

\begin{mythm}
\label{thm:withdrawal_fixed_m}
If there are $n$ TOK outstanding and $\mathcal{D}$ implements a $(\delta, \lambda)$ insurance policy, then when $m$ ETH are illegitimately withdrawn the quantity of ETH held by $\mathcal{D}$ after recovery will be equivalent to
\[
\frac{n^2 - nm + \delta nm}{n + m \delta}.
\]
\end{mythm}

\begin{myproof}
According to Theorem~\ref{thm:sp_price}, absent any exogenous influences, each $\lambda$-SP will increase in value by $x$ in the event that the price of ETH/TOK drops by $x$. Assuming that the withdrawal is seen as being unauthorized, investors will consider $m$ ETH to have been stolen. On the other hand, $\mathcal{D}$ still retains $\delta m$ $\lambda$-SPs and $n-m$ ETH. We seek to determine $\mathcal{E}(m, n, \delta)$, the total ETH remaining in $\mathcal{D}$ after recovery. Let $\Pi(m, n, \delta)$ be the profit of each SP when closed. The two functions are related by the following system of equations.
\begin{equation}
\label{eqn:price_eq}
\begin{array}{rcl}
\mathcal{E}(m, n, \delta) & = & n - m + \delta m ~ \Pi(m, n, \delta) \\
\Pi(m, n, \delta) & = & 1 - \frac{\mathcal{E}(m, n, \delta)}{n} \\
\end{array}
\end{equation}
Solving these equations gives,
\begin{equation}
\begin{array}{rcl}
\mathcal{E}(m, n, \delta) & = & \frac{n^2 - nm + \delta nm}{n + m \delta} \\
\Pi(m, n, \delta) & = & \frac{m}{n + m \delta} \\
\end{array}
\end{equation}
where the expression for $\mathcal{E}(m, n, \delta)$ holds the final result.
\end{myproof}

To develop an intuition for how much ETH can be recovered, we consider several different scenarios. First, suppose that all $n$ ETH were illegitimately withdrawn from the contract. If $\delta = 1$, then $\mathcal{D}$ can restore the ETH holdings to $\sfrac{n}{2}$ or half of the total ether. Increasing $\delta$ to 2 recovers $\sfrac{2n}{3}$ of the ETH. Note that in these scenarios we have recovered a significant portion of ETH even after \emph{all} of the ETH initially in $\mathcal{D}$ (except for the fee $\frac{\delta n}{\lambda}$) was stolen. Any ETH recovered was from closing the $\lambda$-SPs that $\mathcal{D}$ acquired during the withdrawal process. Suppose next that half the ETH is stolen. In this case $\sfrac{2n}{3}$ and $\sfrac{3n}{4}$ ETH remains after closing the $\lambda$-SPs assuming $\delta$ is 1 or 2 respectively. This last scenario raises a question, for a fixed $\delta$, what is the least ETH that will remain in $\mathcal{D}$ after recovery for a given amount of $m$ ETH initially stolen?

\begin{mythm}
\label{thm:mech:eth_remain}
If there are $n$ TOK outstanding and $\mathcal{D}$ implements a $(\delta, \lambda)$ insurance policy, then the least amount of ETH remaining after recovery is $\sfrac{n \delta}{(\delta +1)}$ \emph{regardless of how much ETH is initially stolen}.
\end{mythm}

\begin{myproof}
We begin by finding the derivative of $\mathcal{E}(m, n, \delta)$ with respect to $m$,
\[
\frac{\partial \mathcal{E}}{\partial m} = \frac{-n^2}{(n + \delta m)^2}.
\]
$\frac{\partial \mathcal{E}}{\partial m}$ is an increasing function of $m$. Therefore, $\mathcal{E}$ achieves its maximum value when $m=n$. Substituting $m=n$ back into $\mathcal{E}$ we have the final result.
\end{myproof}

\subsection{Recovery Processor}

The last section discussed the properties of the withdrawal processor, which is designed to help $\mathcal{D}$ hedge against the loss of ETH due to theft. Specifically, $\sfrac{\delta}{\lambda}$ of each withdrawn ETH should be retained for the purpose of opening a $\lambda$-SP. Although we argued that this practice allows for the recovery of a significant portion of stolen ETH, there remains the question of \emph{when} the $\lambda$-SPs should be closed. $\mathcal{D}$ is autonomous and up to this point, the withdrawal mechanism has remained autonomous. Ideally $\mathcal{D}$ will also be capable of deciding when to sell $\lambda$SPs without direct intervention from token holders. On the other hand, it is difficult (if not impossible) to automatically detect theft, even when using the trade price of ETH/TOK as a signal. More significantly, open $\lambda$-SPs are a valuable but volatile asset. The longer a $\lambda$-SP remains open, the more likely it is to receive a margin call if the exchange price ETH/TOK fluctuates by more than $\sfrac{1}{\lambda}$ before the positions are closed.

Consider a simple sale criterion for $\lambda$-SPs based only on the current value of open contracts. Assuming efficient markets, each $\lambda$-SP can rise in value by as much as 1 ETH if the index price drops to 0 (see Section~\ref{sec:back:futures}). The simple approach is to sell each $\lambda$-SP once its value rises above \emph{recovery threshold} $\alpha$ ETH, 

\begin{mythm}
\label{thm:recovery_thresh}
If there are $n$ TOK outstanding and $\mathcal{D}$ implements a $(\delta, \lambda)$ insurance policy with recovery threshold $0<\alpha<\sfrac{1}{(1 + \delta)}$, then there will remain at least 
\begin{equation}
\label{eqn:min_eth_with_alpha}
\min\left[ \frac{n \delta}{(\delta +1)}, ~n\left(1 - \frac{\alpha}{(1-\alpha \delta)}\right) \right]
\end{equation}
ETH after an unauthorized withdrawal and recovery (if it occurs).
\end{mythm}

\begin{myproof}
Theorem~\ref{thm:withdrawal_fixed_m} provides the value of a $\lambda$-SP for the unauthorized withdrawal of $m$ ETH assuming that the $\lambda$-SPs are sold and their ETH added back to $\mathcal{D}$:
\[
\Pi(m, n, \delta) = \frac{m}{n + m \delta}.
\]
By construction, the $\lambda$-SPs are sold when $\Pi(m, n, \delta) > \alpha$. Note that it does not make sense for $\alpha$ to exceed $\sfrac{1}{(1 + \delta)}$ because $\Pi(m, n, \delta) \leq \sfrac{1}{(1 + \delta)}$ for all $m \leq n$. Now assuming $\alpha < \sfrac{1}{(1+\delta)}$, $\Pi(m, n, \delta) > \alpha$ whenever  
\[
\label{eqn:m_thresh}
m > \sfrac{\alpha n}{1- \alpha \delta} = m^\ast.
\]
If $m > m^\ast$ then value then the $\lambda$-SPs are sold and the minimum ETH after recovery is given by Theorem~\ref{thm:mech:eth_remain}: $\sfrac{n \delta}{(\delta +1)}$. But if $m \leq m^\ast$, then no $\lambda$-SPs are sold, so the ETH remaining in $\mathcal{D}$ is as little as 
\begin{equation}
\label{eqn:m_over_thresh}
n - m^\ast = n \left(1 - \frac{\alpha}{(1-\alpha \delta)} \right).
\end{equation}
Since $m$ can be arbitrary, we choose the minimum ETH that can remain for either the case where $m > m^\ast$ or $m \leq m^\ast$. The result follows.
\end{myproof}

To illustrate how the choice of $\alpha$ affects the amount of ETH recovered, suppose that we choose $\alpha = \sfrac{1}{(1+\delta)}$. Then for any choice of $\delta$, Equation~\ref{eqn:m_over_thresh} is equal to 0, which means that Equation~\ref{eqn:min_eth_with_alpha} is also 0. This makes sense because we have set $\alpha$ so high that the $\lambda$-SPs will never be sold, no matter how many ETH are stolen. On the other hand, suppose that we set $\alpha = \sfrac{1}{2(1+\delta)}$ and $\delta = 1$. In this case the left side of Equation~\ref{eqn:min_eth_with_alpha} is $\sfrac{n}{2}$ and the right side is $\sfrac{2n}{3}$, so the minimum ETH recoverable is $\sfrac{n}{2}$.

The left and right sides of Equation~\ref{eqn:min_eth_with_alpha} are equal when $\alpha^{\ast} = \sfrac{1}{(1 + 2 \delta)}$, which makes the minimum recoverable ETH (for any size $m$) equal to $\sfrac{n \delta}{(\delta + 1)}$. As we saw in the example in the previous paragraph, a lower value for $\alpha$ will allow for the recovery of more ETH provided that $m$ is large enough trigger the sale of $\lambda$-SPs. Thus $\alpha^{\ast}$ does not achieve the maximum ETH recoverable for every $m$. However, by choosing $\alpha = \alpha^\ast$, we guarantee that the bound on the minimum ETH recoverable from Theorem~\ref{thm:mech:eth_remain} is achievable for any $m$.

\subsection{Setting Security Parameters}

\begin{figure}[!tbp]
  \centering
  \begin{minipage}[b]{.8\textwidth}
    \includegraphics[width=\textwidth]{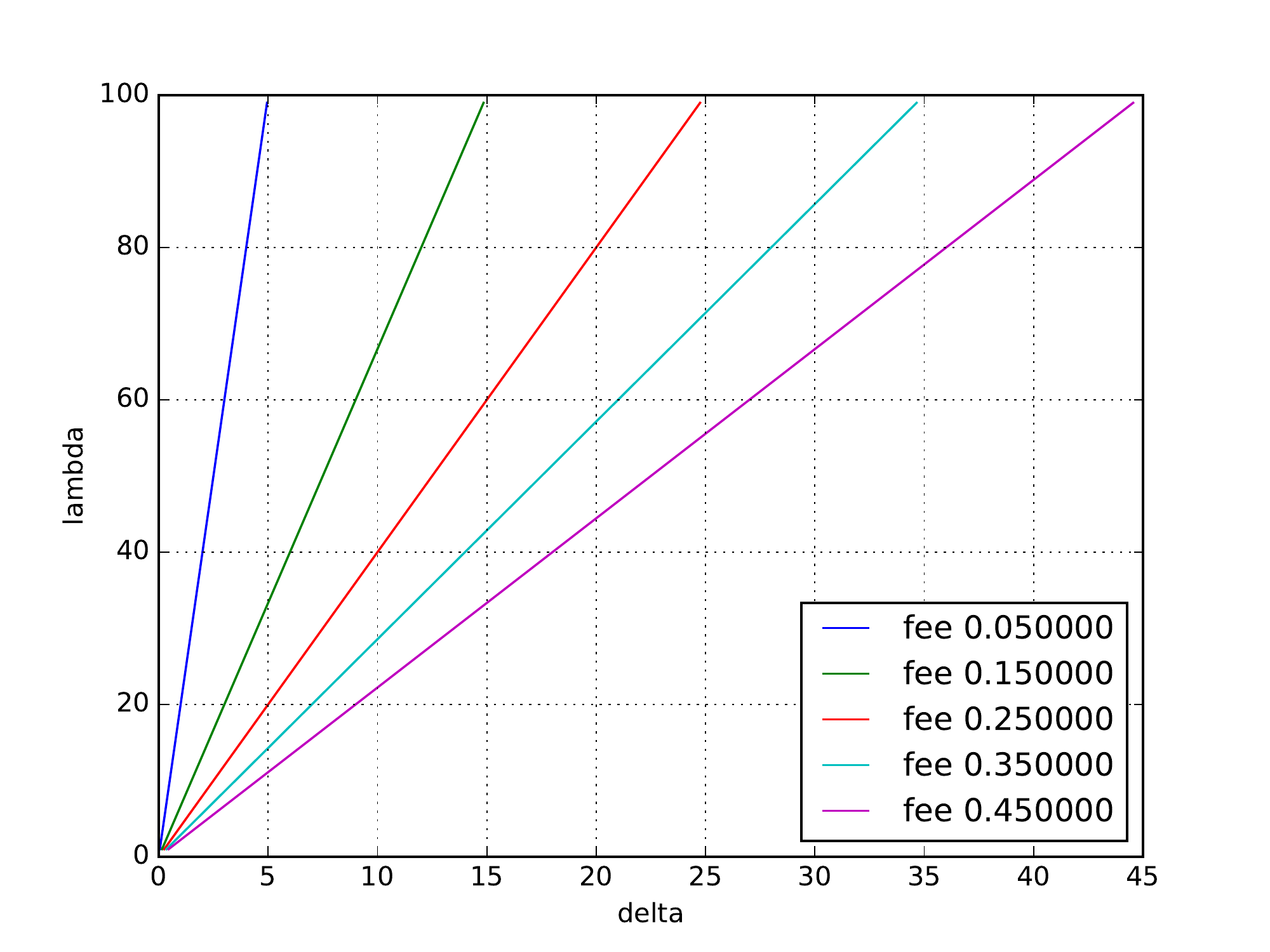}
    \caption{Delta vs. lambda}
    \label{fig:delta_lambda}
  \end{minipage}
  \end{figure}

  \begin{figure}[!tbp]
    \centering
  \begin{minipage}[b]{0.8\textwidth}
    \includegraphics[width=\textwidth]{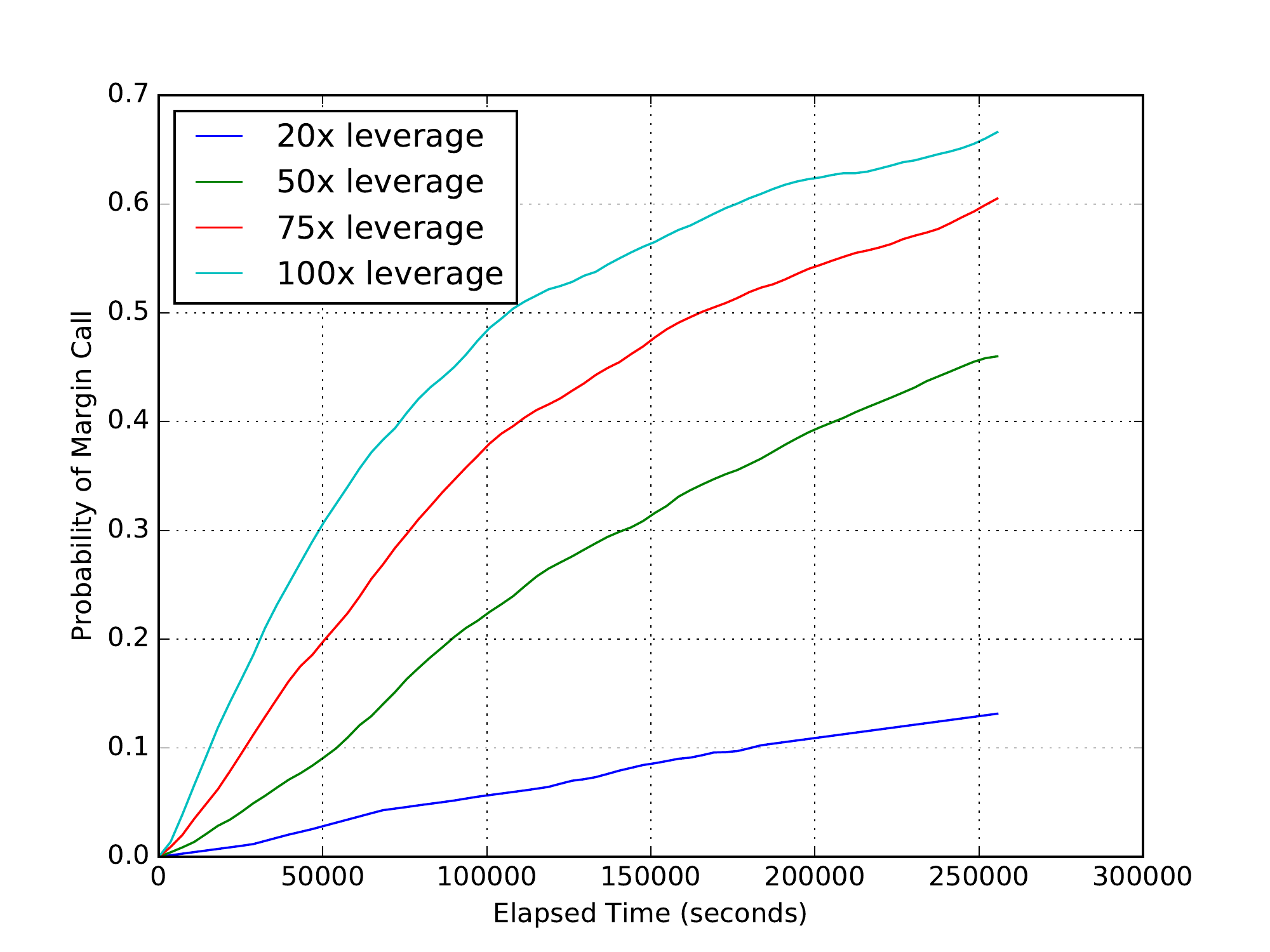}
    \caption{Probability of margin call.}
    \label{fig:lambda_margin}
  \end{minipage}
\end{figure}
Two parameters controlled by $\mathcal{D}$ are critical to the security of the DAPP withdrawal mechanism: $\lambda$, the SP leverage and $\delta$, the number of $\lambda$-SPs purchased for each ETH withdrawn. Setting these parameters presents a tradeoff in terms of security and usability. The risk of margin call decreases with $\lambda$, however it also raises the withdrawal fee. And the minimum ETH recoverable goes up with $\delta$, but it too raises the withdrawal fee. Figure~\ref{fig:delta_lambda} shows the tradeoff between $\lambda$ and $\delta$ with each line representing a fixed withdrawal fee. For example, in order to maintain a 5\% withdrawal fee (blue line) $\lambda$ can be as low as 20, but $\delta$ is limited to 1. Thus according to Theorem~\ref{thm:mech:eth_remain}, the minimum ETH recoverable is only $\sfrac{1}{2}$. However, when the withdrawal fee is 25\% (red line), the leverage can remain 20 with $\delta$ increasing to 5. Now the minimum ETH recoverable is $\sfrac{5}{6}$, more than 80\%.

It is difficult to predict the probability of a margin call for $\lambda$-SPs without knowing the dynamics of the futures market or the volatility of the underlying index price of ETH/TOK. To get a better idea of the price dynamics of a real cryptocurrency, we analyzed one month of Bitcoin futures data from the OkCoin exchange from August 5 until September 4, 2016. For various values of $\lambda$ Figure~\ref{fig:lambda_margin} shows the empirical probability that a $\lambda$-SP will receive a margin call after a given period of time. After a 3 days, the probability of a margin call is less than 15\% when $\lambda = 20$, but when $\lambda = 100$ it rises to nearly 70\%. Bitcoin is the oldest and most popular cryptocurrency, thus we expect that its price volatility will be lower than a newly introduced DAPP token. Accordingly, the estimated probabilities of margin call given in Figure~\ref{fig:lambda_margin} should be interpreted as lower bounds on the corresponding probabilities for TOK.

\section{Conclusion}

We have presented a generic mechanism that can be used to insure the ether holdings of DAPPs similar to The DAO. Our approach is to charge a small withdrawal fee, which is used to purchase futures contracts that hedge against a drop in the price of tokens issued by the DAPP. We show that, even in the event that all ether holdings are stolen from the DAPP, the majority of the ether can typically be restored with high probability. More specifically, we find that in order to ensure that a large fraction of assets are recoverable it is necessary that either the volatility of futures contracts remains relatively low or the withdrawal fee remains high. 

\bibliographystyle{splncs03}
\bibliography{references}

\begin{thebibliography}{10}
\providecommand{\url}[1]{\texttt{#1}}
\providecommand{\urlprefix}{URL }

\bibitem{bitmex:2016}
Bitmex. \url{https://www.bitmex.com} (November 2016)

\bibitem{etheria:2016}
Etheria. \url{http://etheria.world} (November 2016)

\bibitem{Gnosis:2016}
Gnosis. \url{https://www.gnosis.pm} (November 2016)

\bibitem{okcoin:2016}
{OkCoin}. \url{https://www.okcoin.com/} (November 2016)

\bibitem{oraclize:2016}
Oraclize.it. \url{http://www.oraclize.it} (November 2016)

\bibitem{rootstock:2016}
{Rootstock}. \url{http://www.rsk.co} (November 2016)

\bibitem{SafeMarket:2016}
Safemarket. \url{https://safemarket.github.io} (November 2016)

\bibitem{slock:2016}
{Slock.it}. \url{https://slock.it} (November 2016)

\bibitem{bitcoin:contracts}
{Bitcoin Smart Contracts}. \url{https://en.bitcoin.it/wiki/Contract} (January
  2017)

\bibitem{Back:2014}
Back, A., Corallo, M., Dashjr, L., Mark, F., Maxwell, G., Miller, A., Poelstra,
  A., Tim\'{o}n, J., Wuille, P.: {Enabling Blockchain Innovations with Pegged
  Sidechains}. \url{http://www. opensciencereview.
  com/papers/123/enablingblockchain-innovations-with-pegged-sidechains}
  (October 2014)

\bibitem{Buterin:2017}
Buterin, V.: {Thinking About Smart Contract Security}.
  \url{https://blog.ethereum.org/2016/06/19/thinking-smart-contract-security}
  (January 2017)

\bibitem{Castillo:2016a}
del Castillo, M.: {Ethereum Executes Blockchain Hard Fork to Return DAO Funds}.
  \url{http://www.coindesk.com/ethereum-executes-blockchain-hard-fork-return-dao-investor-funds}
  (July 2016)

\bibitem{Castillo:2016}
del Castillo, M.: {The DAO Attacked: Code Issue Leads to \$60 Million Ether
  Theft}.
  \url{http://www.coindesk.com/dao-attacked-code-issue-leads-60-million-ether-theft}
  (June 2016)

\bibitem{Douceur:2002}
Douceur, J.: {The {Sybil} Attack}. In: Proc. Intl Wkshp on Peer-to-Peer Systems
  (IPTPS) (Mar 2002)

\bibitem{ethereum}
{Ethereum Homestead Documentation}. \url{http://ethdocs.org/en/latest/}

\bibitem{Fischer:1985}
Fischer, M.J., Lynch, N.A., Paterson, M.S.: Impossibility of distributed
  consensus with one faulty process. Journal of the ACM (JACM)  32(2),
  374--382 (1985)

\bibitem{Nakamoto:2009}
Nakamoto, S.: {Bitcoin: A Peer-to-Peer Electronic Cash System}.
  \url{https://bitcoin.org/bitcoin.pdf} (May 2009)

\bibitem{Siegel:2016}
Siegel, D.: {Understanding The DAO Attack}.
  \url{http://www.coindesk.com/understanding-dao-hack-journalists} (June 2016)

\bibitem{Torpey:2016}
Torpey, K.: {Millions of Dollars Worth of ETC May Soon Be Dumped on the
  Market}.
  \url{https://bitcoinmagazine.com/articles/millions-of-dollars-worth-of-etc-may-soon-be-dumped-on-the-market-1472567361}
  (August 2016)

\end{thebibliography}
\end{document}